\documentclass{PoS}

\usepackage{amssymb}
\usepackage{amsmath}
\usepackage{xspace}
\usepackage[labelfont=bf,font=small]{caption}
\usepackage{slashed}

\usepackage{microtype}
\usepackage{siunitx}
\usepackage{cleveref}
\usepackage{nicefrac}
\usepackage[utf8]{inputenc}

\DeclareMathAlphabet{\mathcal}{OMS}{cmsy}{m}{n}

\title{Anomalous couplings in WZ production beyond NLO QCD}
\ShortTitle{Anomalous couplings in WZ production beyond NLO QCD}

\author{\speaker{Robin~Roth}\footnote{ KA-TP-32-2016 }\\
  Institute for Theoretical Physics, KIT, 76128 Karlsruhe, Germany \\
  E-mail: \email{robin.roth@kit.edu}}

\author{Francisco~Campanario\\
  Theory Division, IFIC, University of Valencia-CSIC, E-46980 Paterna, Valencia, Spain\\
  E-mail: \email{francisco.campanario@ific.uv.es}}
\author{Sebastian~Sapeta\\
  CERN PH-TH, CH-1211, Geneva 23, Switzerland\\
  E-mail: \email{sebastian.sapeta@cern.ch}}
\author{Dieter~Zeppenfeld\\
  Institute for Theoretical Physics, KIT, 76128 Karlsruhe, Germany\\
  E-mail: \email{dieter.zeppenfeld@kit.edu}}

\abstract{
We study WZ production with anomalous couplings at \nNLO QCD using
the LoopSim method in combination with the Monte Carlo program \VBFNLO.
Higher order corrections to WZ production are dominated by additional hard jet radiation.
Those contributions are insensitive to anomalous couplings and should thus be removed in analyses.
We do this using a dynamical jet veto based on the transverse energy of the QCD and EW final
state particles.
This removes jet dominated events without introducing problematic logs like a fixed $\pt$ jet
veto.}

\FullConference{Fourth Annual Large Hadron Collider Physics\\
		13-18 June 2016\\
		Lund, Sweden}

\newcommand{\als}{\alpha_{\rm{s}}}

\newcommand{\mur}{\mu_{\text{R}}}
\newcommand{\muf}{\mu_{\text{F}}}

\newcommand{\nNLO}{\nbar \text{NLO}\xspace} 

\newcommand{\nbar}{\ensuremath{\bar{n}}}

\newcommand{\pt}{p_{\text{T}}{}}
\newcommand{\et}{\text{E}_{\text{T}}}
\newcommand{\etj}{\sum_{\text{jets}}\text{E}_{\text{T},i}}
\newcommand{\etew}{\text{E}_{\text{T},W} + \text{E}_{\text{T},Z}}

\newcommand{\ptj}{\pt_{\text{jet}}}

\newcommand{\ptz}{\pt_{\text{Z}}}

\newcommand{\ptlmax}{\pt_{\text{l,max}}}

\newcommand{\htjets}{H_{\text{T,jets}}\xspace}
\newcommand{\HT}{H_{\text{T}}\xspace}
\newcommand{\HTval}{ \frac{1}{2} \left(\sum_{\text{partons}} p_{T,i} 
    + \sum_{W,Z}\sqrt{p_{T,i}^2+m_{i}^2}\right) \xspace }

\newcommand{\VBFNLO}{{\tt VBFNLO}\xspace}

\newcommand{\xjetval}{\frac{\etj}{\etj+\etew}}
\newcommand{\xjet}{x_{\text{jet}}}

\newcommand{\MSM}{\mathcal{M}_{\text{SM}}}
\newcommand{\MAC}{\mathcal{M}_{\text{AC}}}
\newcommand{\MACs}{\mathcal{M}^{d=6}_{\text{AC}}}
\newcommand{\MACe}{\mathcal{M}^{d=8}_{\text{AC}}}

\newcommand{\ow}{\mathcal{O}_W}
\newcommand{\owval}{\left(D_\mu \Phi \right)^\dag \widehat{W}^{\mu\nu} \left(D_\nu \Phi \right)}
\newcommand{\ob}{\mathcal{O}_B}
\newcommand{\obval}{\left(D_\mu \Phi \right)^\dag \widehat{B}^{\mu\nu} \left(D_\nu \Phi \right)}

\newcommand{\owww}{\mathcal{O}_{WWW}}
\newcommand{\owwwval}{\rm{Tr}\left[ \widehat{W}_{\mu\nu} \widehat{W}^{\nu\rho} \widehat{W}^\mu_\rho \right]}

\begin{document}

\section{Introduction}

The production of two heavy vector bosons is an interesting process, as it allows to study the
interactions between the gauge bosons.
Leptonic decays of the vector bosons can be used to discriminate backgrounds and get precise
measurements of the kinematics.
Searches for new physics include new resonances decaying to vector boson pairs and changes to
their couplings due to new particles in loops.
Those can appear in tails of distributions or as modifications in angular distributions
sensitive to spin.
To measure diboson production and search for deviations from the Standard Model (SM) high center of mass energies and high luminosities are required,
such that the most recent LHC upgrades promise to enhance those analyses.

In this contribution, we study WZ production as a representative diboson production process.
We calculate NLO QCD corrections to WZ and WZj production with VBFNLO and combine them to get \nNLO accuracy using the
LoopSim approach \cite{Rubin:2010xp}.
This improves the prediction especially for high-$\pt$ vector bosons.
Recently a calculation of the inclusive cross section of WZ production at NNLO QCD was
reported in Ref.~\cite{Grazzini:2016swo}, but there are no differential distributions
available yet.
\footnote{On NNLO QCD calculations of diboson production see the contribution by
  S.~Kallweit, ``NNLO di-boson production.'' }

We study Anomalous Couplings (AC) using an Effective Field Theory (EFT) approach.
This allows to parametrize deviations from the Standard Model (SM) interactions, especially
in triple and quartic gauge couplings in a general way without choosing a specific model.
Effects of AC are most visible in the region of high invariant-mass diboson pairs with high transverse momentum of the final state particles.
In particular in the high-$\pt$ region, the LoopSim approach should approximate the full NNLO corrections closely.

WZ production has been measured at the LHC in several decay channels. 
The fully leptonic decay, as studied in
Ref.~\cite{Aad:2012twa,Aad:2014pha,Aad:2016ett,Aaboud:2016yus,Khachatryan:2016tgp},
has little background,
but at the same time the smallest cross section.
Semi-leptonic final states, \cite{Chatrchyan:2012bd,Aad:2013pdy},
 have a larger cross sections but suffer from backgrounds due to top
production and single-vector-boson + jets. 
Measuring the cross section allows to set limits on anomalous triple gauge couplings,
as many of the analyses do.
For details on diboson measurements see also the contributions by S.L.~Barnes and N.~Woods
in these proceedings.

\section{Calculational setup}
For the simulation, we use \VBFNLO \cite{Arnold:2008rz,Baglio:2014uba,Arnold:2011wj}
in combination with LoopSim \cite{Rubin:2010xp}.
The setup is comparable to previous work with LoopSim on WZ \cite{Campanario:2012fk}, WW \cite{Campanario:2013wta} and ZZ production \cite{Campanario:2015nha}.
LoopSim generates a merged sample of WZ@NLO and WZj@NLO \cite{Campanario:2010hp}
to give us WZ@$\nNLO$.
For the study of AC, we are interested in phase space regions with 
high transverse momentum and high invariant mass of electroweak particles.
These can get large contributions from additional QCD radiation of 
$\mathcal{O}( \als \ln^2 \ptj / m_Z )$, which are included in LoopSim.
Compared to the full NNLO calculation, LoopSim misses the finite 2-loop contributions.
Those are expected to be on the level of a few percent, since they are suppressed by $\als^2 $
and are not enhanced, like some real emission contributions.
Due to the missing contributions the prediction of the total cross section, the low-$\pt$
phase space region and the scale dependence is only NLO.

\subsection{Anomalous Couplings} 
To parametrize AC, we use the set of dimension-6 operators based on the HISZ basis
\cite{Hagiwara:1993ck}
as updated in Ref.~\cite{Hankele:2006ma}. 
They extend the SM via 
$\mathcal{L} = \mathcal{L_{\text{SM}}} + \sum_i \frac{f_i}{\Lambda^2}\mathcal{O}_i$.
Assuming C and P invariance, only three operators contribute to the WWZ vertex:
\begin{align}
  \ow &= \owval, \\
  \ob &= \obval, \\
  \owww &= \owwwval .
  \label{eq:wwz-operators}
\end{align}
We focus on the $\ow$ operator for this work.
It is a suitable representative for AC as it leads to a term with non-SM Lorentz structure in the WWZ
vertex.
To get meaningful limits on AC, one has to consider a complete set of operators, which
affect a given observable, including correlations between operators.

Diboson production processes are competitive in limiting dimension 6 operators and are
therefore included in global fits of AC limits, like
Refs.~\cite{Corbett:2012ja,Masso:2012eq,Butter:2016cvz}. 
Our values used for the couplings are in the allowed region of Ref.~\cite{Corbett:2012ja} and represent typical values current
measurements are sensitive to.

\subsection{Validity of EFT approach}
\label{sec:acinterfer}
Effective Field Theory (EFT) for triple gauge couplings assumes an expansion in
$\frac{f}{\Lambda^2}$.  
This depends on both the coupling of new physics as well as its energy scale.
The expansion is only valid if the scale of the considered observable is smaller than the new physics scale entering $\Lambda$. 

For high invariant masses, the EFT operators violate unitarity of the S-matrix. 
This is a sign that the EFT approach is no longer valid and should be replaced by a UV
complete model.
To still get physical predications, one can apply a unitarization procedure.
We use the form factor
\begin{align}
	FF = \left( 1+\frac{m_{\text{WZ}}}{\Lambda_{\text{FF}}^2} \right)^{\text{-n}}, \quad
  \text{with } \Lambda_{\text{FF}} = \SI{2}{TeV} \text{ and n} = 1 .
	\label{eq:ffdef}
\end{align}
The value for the form factor scale is determined with the \VBFNLO form factor tool
\cite{VBFNLOFF:2014}, such that the operator does not lead to unitarity violation in
$2\rightarrow 2$ scattering.
The exponent n is chosen to cancel the leading divergence of the EFT
operator. Using the form factor tool the scale $\Lambda_{\text{FF}}$ is fixed, such that
this description leaves no free parameter in the unitarization scheme.
For most plots, we will focus on phase space regions significantly below $\Lambda_\text{FF}$
where neither the form factor nor most unitarization procedures would have a
visible effect.

Even below the scale of unitarity violation, there is an ambiguity in how to
make predictions using an EFT.
In the calculation of the squared matrix element $\left|\mathcal{M}\right|^2$, there are
interference terms between the SM and AC as well as purely AC terms.
If one considers an amplitude with contributions from AC operators of dimension 6 and
dimension 8, the terms are:
\begin{align}
  \mathcal{M} &= \MSM + \underbrace{\MACs}_{\nicefrac{1}{\Lambda^2}} +
  \underbrace{\MACe}_{\nicefrac{1}{\Lambda^4}} +\mathcal{O}(\Lambda^{-6}) \\
  \left|\mathcal{M}\right|^2 &= 
  \underbrace{\left|\MSM\right|^2}_{\nicefrac{1}{\Lambda^0}} 
  + \underbrace{2 \text{Re} \MSM^* {\MACs}}_{\nicefrac{1}{\Lambda^2}}
  + \underbrace{\left| \MACs\right|^2 }_{\nicefrac{1}{\Lambda^4}}
  + \underbrace{2 \text{Re} \MSM^* {\MACe}}_{\nicefrac{1}{\Lambda^4}}
  + \underbrace{\left|\MACe\right|^2}_{\nicefrac{1}{\Lambda^8}}
  +\mathcal{O}(\Lambda^{-6})
\end{align}

We include both the $ \MSM^* \MACs$ and the ${\left|\MACs\right|}^2$ term in our calculation.
By naive power counting, one would assume that the latter should be considered
simultaneously with dim-8 operators, that contribute via $\MSM^* \MACe$.
This is in general not the case because the SM amplitude is suppressed by the weak coupling, such that
$\left| \MACs \right|^2$ can naturally be larger than $\MSM^* \MACe$.

Besides this size argument, there are also practical reasons to include the 
$\left| \MAC \right|^2$ term. 
Without it, one can generate (unphysical) negative cross
sections when negative interference exceeds the SM contribution.

To be independent of this ambiguity, one can restrict oneself to phase space regions where
the squared term is not relevant. Based on the sign dependence of the interference, we will
study in which phase space regions both terms contribute in \cref{sec:acnnlo}.

\section{Numerical results}

We consider the LHC at run 2 with pp collisions at \SI{13}{TeV}.  
The jets are clustered using the anti-$k_t$
algorithm~\cite{Cacciari:2008gp} with a cone radius of $R=0.4$. To simulate
typical detector acceptance, we impose a minimal set of inclusive
cuts
\begin{equation}
\label{eq:basiccuts}
\begin{aligned}
		% \pt_l &> \SI{15}{GeV} & \pt_{j} &>\SI{30}{GeV} & \slashed E_{\text{T}} &> \SI{30}{GeV} \\
		\pt_l &> 15 \text{GeV} & \pt_{j} &> 30 \text{GeV} & \slashed E_{\text{T}} &> 30 \text{GeV} \\
		|y_j| &< 4.5  & |\eta_l| &< 2.5 &  R_{lj} &> 0.4 \\
	 120 \text{GeV} &> m_{ll} > 60 \text{GeV} ,
\end{aligned}
\end{equation}
where the $m_{ll}$ cut is applied only to same-flavor leptons with opposite sign coming from the $Z$ boson.

We consider decays $W\rightarrow e \nu_e, Z\rightarrow \mu \mu$.
Adding the other leptonic final states increases the number of expected events by a factor
of $4$.\footnote{The factor 4 is not exact, because of small corrections due to the
  not-ideal reconstruction in states with identical flavors and the Pauli interference effect.}

For the renormalization scale $\mur$ and factorization scale $\muf$, we use 
\begin{equation}
  \mu_0 = \HT = \HTval .
\end{equation}

The theoretical uncertainty is estimated using a simultaneous variation of the two scales  by
a factor of 2: $\mur=\muf=\{0.5,2\}\mu_0$.
We use the default values set in VBFNLO 3.0 for electroweak constants and NNPDF23 \cite{Ball:2012cx}
for the parton distribution functions.

\subsection{WZ production at \nNLO QCD}
\begin{figure*}[tp]
	\centering
	\includegraphics[width=.45\textwidth]{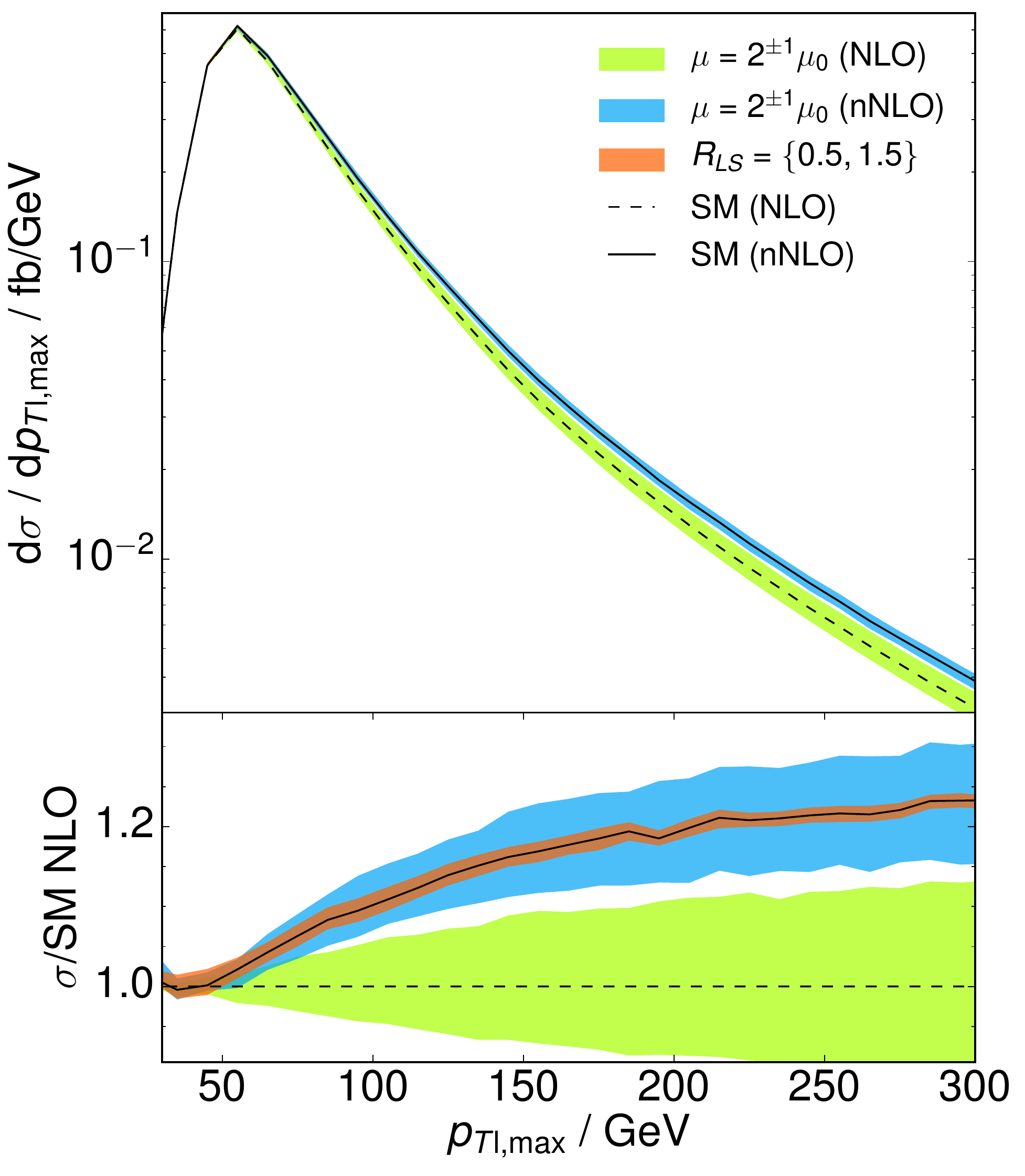}
	\includegraphics[width=.45\textwidth]{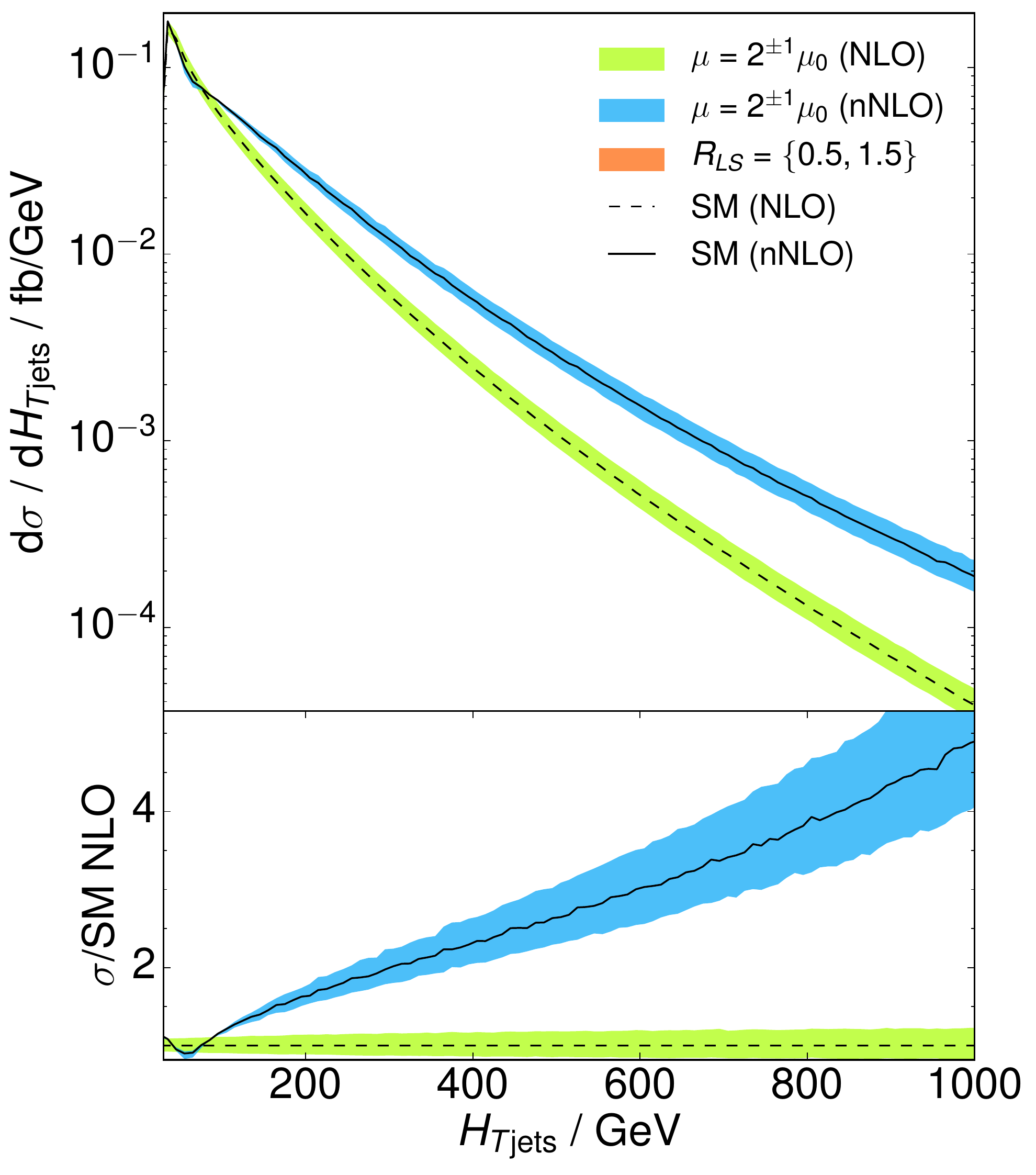}
  \caption{The $\pt$ of the leading lepton and $\htjets$ are shown for WZ production at NLO
    (dashed) and \nNLO QCD (solid). 
    Corrections are typically below 25\% as seen on the left for $\ptlmax$, but can be huge in observables sensitive to additional emissions,
    for example they reach a factor of 5 in $\htjets$ at \SI{1}{TeV}.
    The scale variation is given as a band for NLO (green) and \nNLO (blue).
    For \nNLO, also the variation of the LoopSim clustering parameter $\rm{R}_{\text{LS}}$ is
    indicated as a band.
    Its influence is small for most observables and highlights phase space regions sensitive
    to details of the clustering.
  }
\label{fig:loopsim_large_kfac}
\end{figure*}

Results for WZ production at \nNLO QCD using Loopsim were first presented in
Ref.~\cite{Campanario:2012fk}.
As shown there, the corrections due to \nNLO on top of NLO QCD can be sizeable.
A typical electroweak observable, like the $\pt$ of the hardest lepton ($\ptlmax$), is
enhanced by about 25\% at \SI{300}{GeV}. 
Changes are substantially larger for observables sensitive to extra radiation.
For $\htjets$, the correction is a factor of 5 at \SI{1}{TeV}.
These distributions are shown in \cref{fig:loopsim_large_kfac}.
In both cases, the corrections are outside of the scale variation.

\subsection{AC at \nNLO QCD}
\label{sec:acnnlo}
For AC, we include both the $\MSM^* \MAC$ interference term as well as the $\left|\MAC\right|^2$ term, as
discussed in \cref{sec:acinterfer}.
The interference term is sensitive to the sign of the AC, while the squared term is not.
Which of the two terms dominated can thus be seen in distributions by looking for 
dependence on the coupling sign.

\begin{figure*}[tp]
	\centering
	\includegraphics[width=0.48\textwidth]{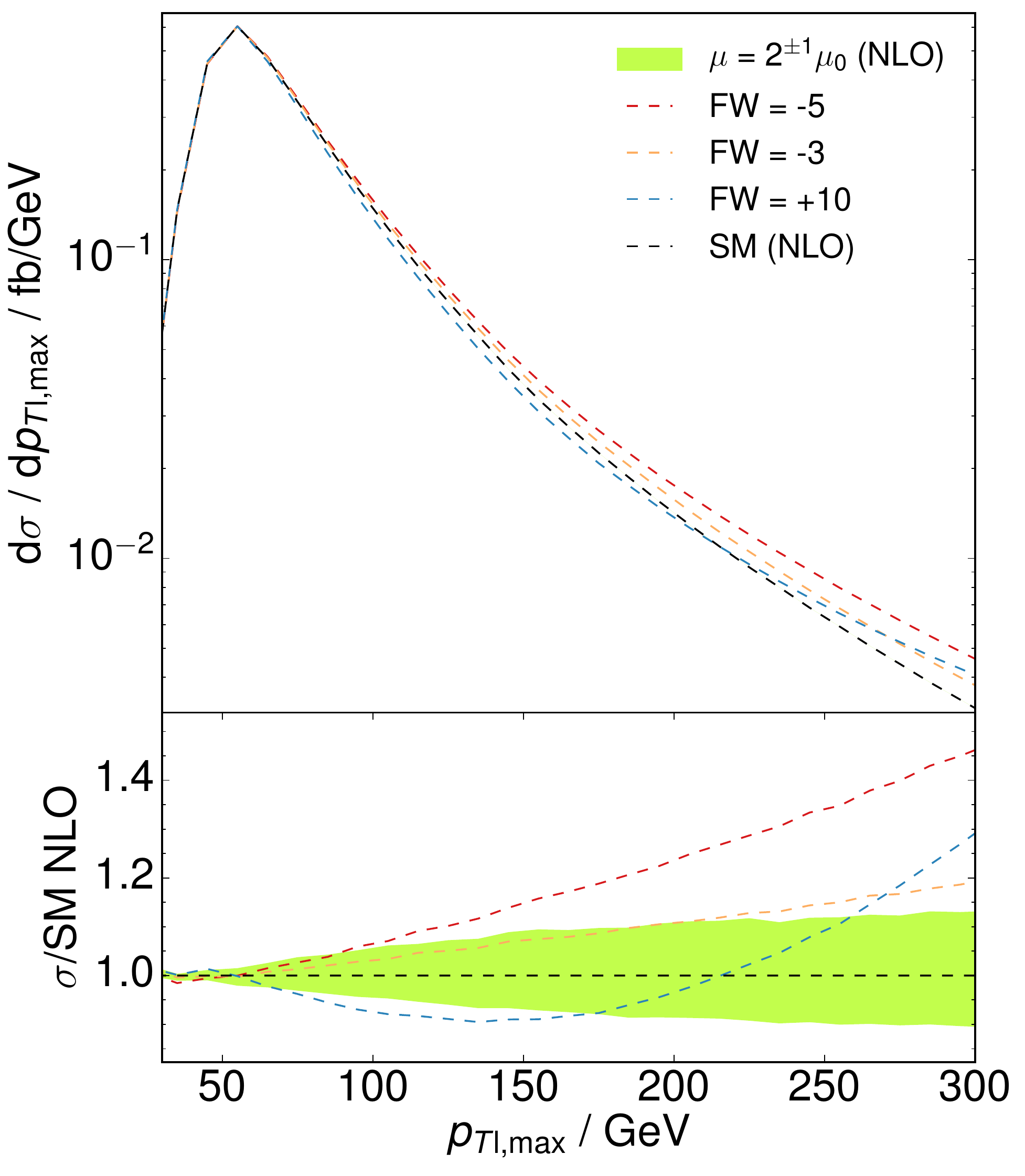}
	\includegraphics[width=0.48\textwidth]{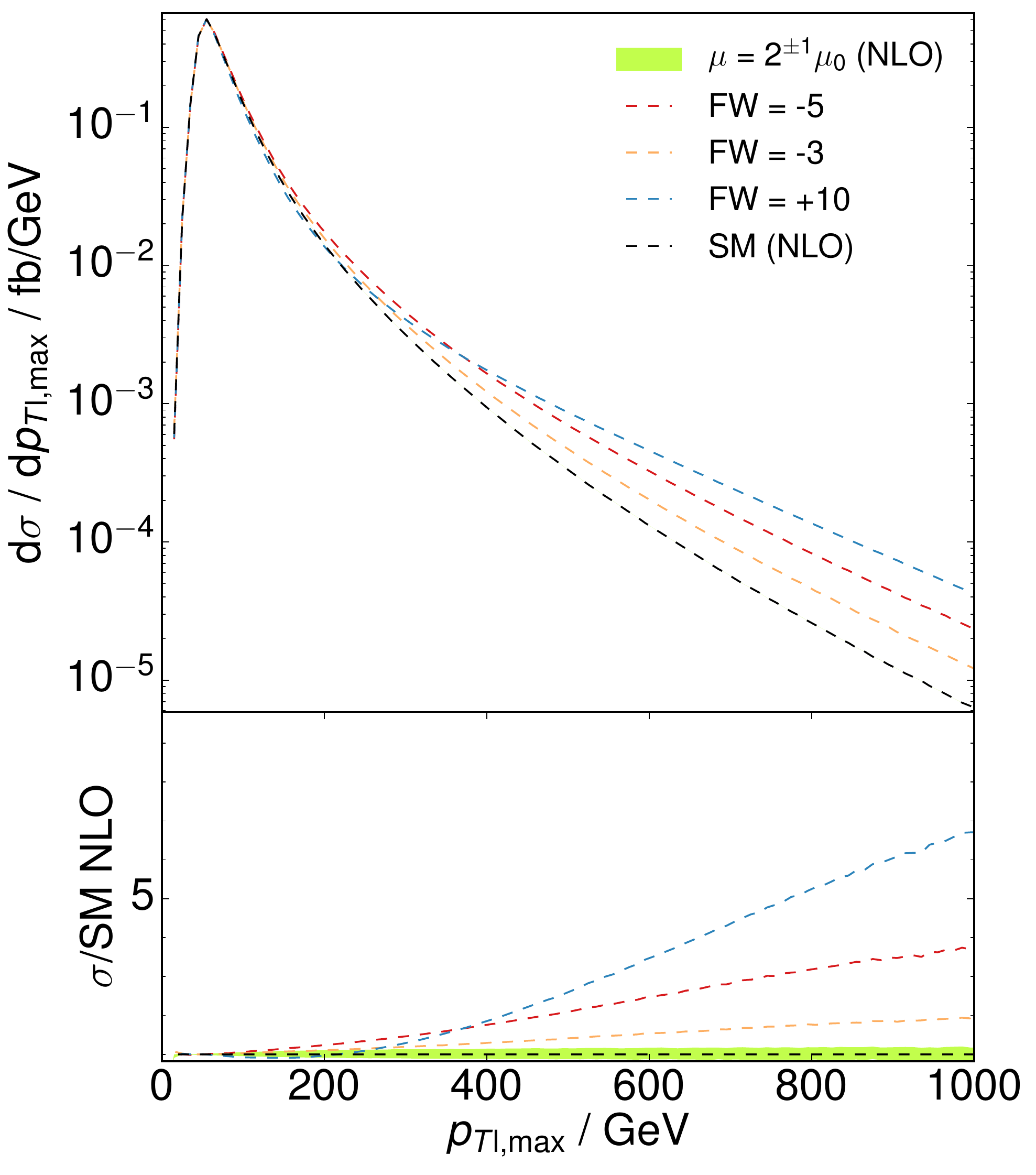} 
  \\
	\caption{The effect of the AC operator $\ow$ is shown on the $\pt$ of the leading lepton
    at NLO QCD for different regions of $\pt$.
    The effect grows with $\pt$, such that searches for AC are mostly focused in the
    high-$\pt$ bins of distributions where an enhancement of the SM cross section by a
    factor of 5 or more is possible (right).
    The chosen AC values have interferences with the SM most prominent at \SI{150}{GeV} (left).
  }
	\label{fig:motivation_highptl}
\end{figure*}

To show the typical effect of AC, we consider $\ptlmax$ at NLO QCD in
\cref{fig:motivation_highptl}.
For $\ptlmax$ and the chosen AC values, at around \SI{150}{GeV}, the destructive
interference is maximal, while above \SI{300}{GeV}, the $\left|\MAC\right|^2$ term dominates.
The interference region gives access to the sign.

\begin{figure*}[tp]
	\centering
	\includegraphics[width=0.45\textwidth]{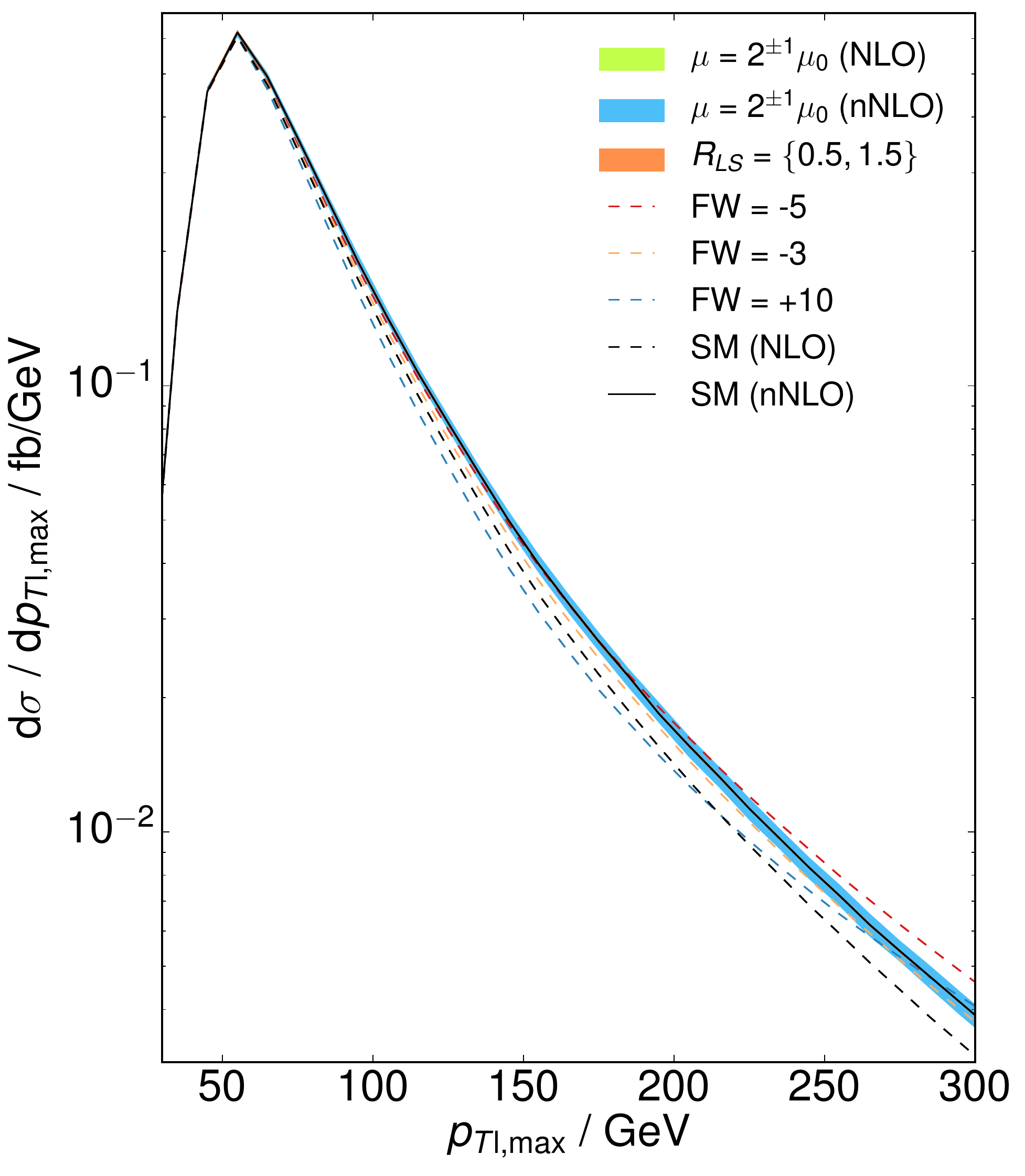}
	\includegraphics[width=0.45\textwidth]{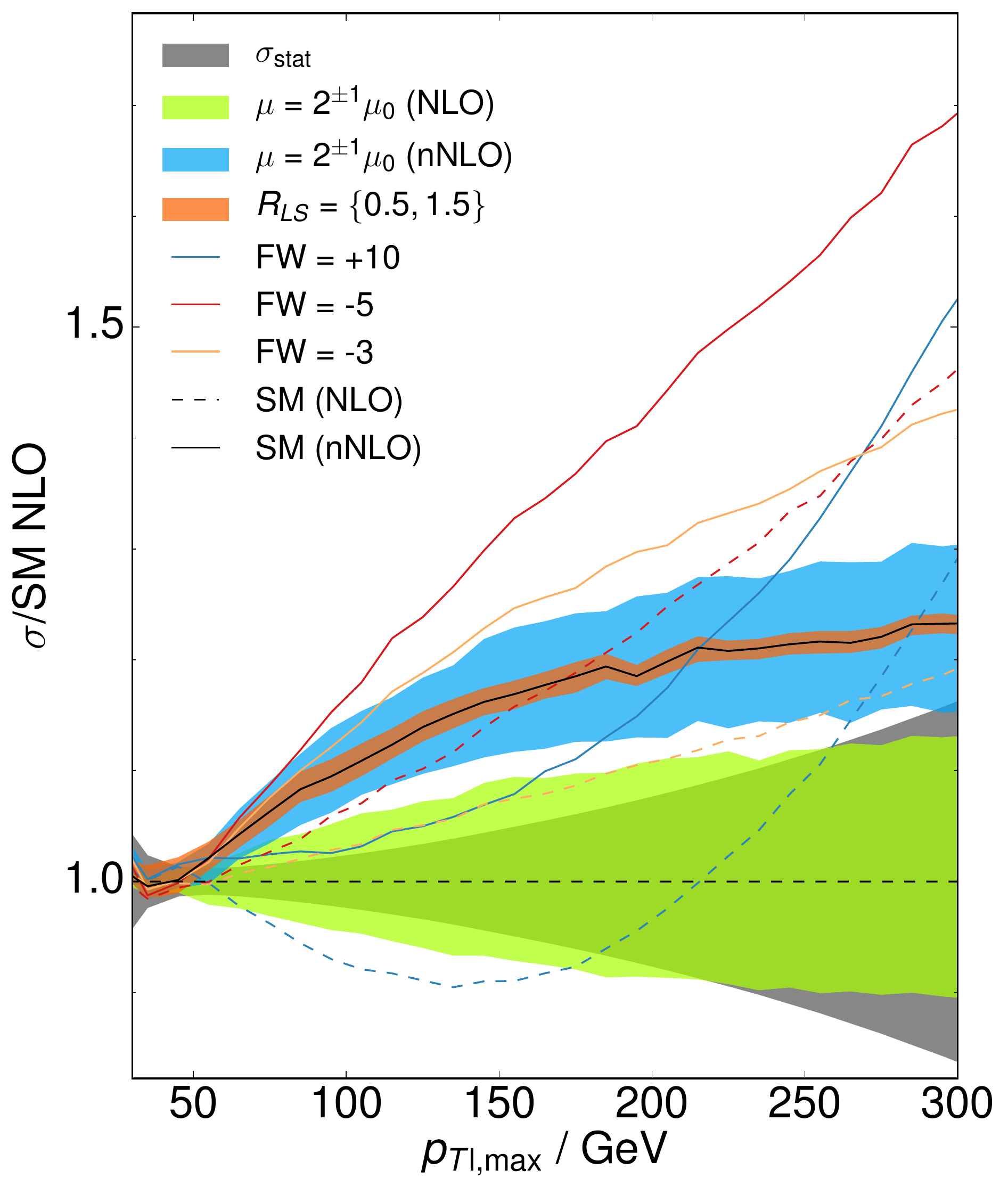}\\
	\caption{On the left the transverse momentum of the hardest lepton is shown.
The right shows the same observable, normalized to the SM prediction at NLO QCD.
As shown in \cref{fig:loopsim_large_kfac}, the $\nNLO$ corrections are on the order of
25\%.
Non-vanishing AC can change the cross section by the same amount. 
Colored dashed lines show the distribution for AC at NLO QCD, which (depending on the
bins considered) give the same value as the SM $\nNLO$ prediction.
For a AC analysis, also all QCD corrections have to be taken into account, leading to the
solid colored lines for AC at $\nNLO$ QCD.  }
	\label{fig:confuse_nnlo_ac}
\end{figure*}

\begin{figure}[tp]
  \centering
  \includegraphics[width=0.45\textwidth]{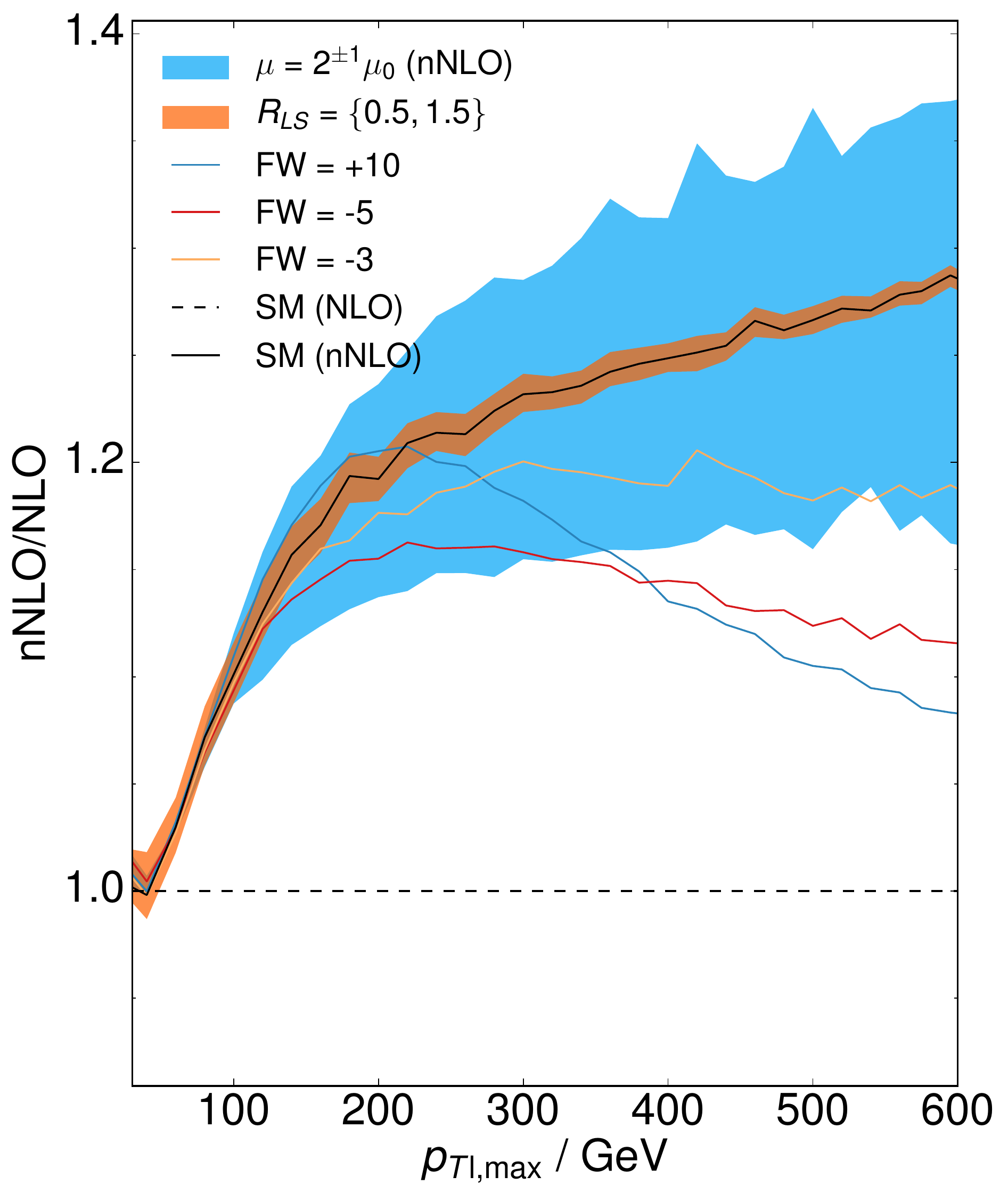}
  \caption{The ratio $\left( \rm{d}\sigma_{\nNLO}/\rm{d}\sigma_{\text{NLO}} \right)_{ 
      \text{SM},\text{AC} } $
    for
    $\ptlmax$ is shown.
    This K-factor depends significantly on the value of the anomalous coupling, ranging from 1.1 to 1.3
    for the bin at \SI{600}{GeV}.
    Therefore, AC can not be accurately described by rescaling an AC prediction at NLO with a SM
    $\nNLO$/NLO K-factor.
  }
  \label{fig:ptlmax-nnlo-kfac}
\end{figure}

The corrections at $\nNLO$ shown in \cref{fig:loopsim_large_kfac} are of comparable size to those
due to AC.
Therefore, an analysis based on a prediction at NLO QCD for the SM might mistake a deviation
for a detection of AC, while a prediction at higher order in QCD might match the measurement.
\Cref{fig:confuse_nnlo_ac} shows a comparison of SM $\nNLO$ and AC NLO that are of similar
size.
With \VBFNLO and LoopSim, predictions for AC can also be made at $\nNLO$ accuracy.
The $\nNLO/\text{NLO}$ K-factor depends on the specific value of the anomalous coupling, such
that extrapolating the SM K-factor to AC predictions gives inaccurate results.
This can be seen in \cref{fig:ptlmax-nnlo-kfac}.

\subsection{Dynamical jet veto}
In Ref.~\cite{Campanario:2014lza}, we suggested a dynamical jet veto to improve the
sensitivity to AC in WZj production. We use this veto and study its effect on LoopSim corrections.
AC effects typically grow with the invariant mass or momentum transfer at triple gauge vertices. 
To enhance their signal, the focus is on high-invariant mass boson pairs and high transverse momentum
bosons/leptons in the final state.
When a high-$\pt$ vector boson is required, this occurs about half of the time due to
recoil against a jet (instead of the second vector boson).
To reduce those events, one introduces a jet veto.
A traditional fixed-$\pt$ jet veto rejects all events with additional jets above a certain
threshold.
This introduces logarithms of the veto scale, which need to be resummed.
Also this veto cuts away relevant phase space, since in very-high invariant-mass regions
(for example $m_{WZ} = \SI{1}{TeV}$) most events will have additional jet radiation at
\SIrange{50}{100}{GeV}, which is soft compared to the EW system and does not reduce the AC
sensitivity. Thus, a fixed veto at \SI{50}{GeV} will remove relevant signal in that phase space
region. 

\setlength{\abovedisplayskip}{5pt}
\setlength{\belowdisplayskip}{5pt}

The dynamical veto is based on
\begin{equation}
	\xjet = \xjetval, \text{ where }
  \et = E \frac{\left|\vec p_{\rm{t}}\right|}{\left| \vec p \right|} .
	\label{eq:xjetdef}
\end{equation}
% \vspace{-0.5ex}
The definition of $\et$ here differs from the one chosen in Ref.~\cite{Campanario:2014lza}, 
where $m_T = \sqrt{m^2 + \pt^2}$ was used instead.
At small $\pt$, the latter is dominated by the mass and
thus leads to small $\xjet$ values for inclusive samples, while $\et$ generates a broader
distribution also in the low-$\pt$ region and leads to better discrimination between SM and
AC contributions.
In the high-$\pt$ region or for massless particles, these definitions become identical: $m_{\rm{T}} = \et = \pt$.

\begin{figure}[tbp]
  \begin{center}
    \includegraphics[width=0.45\textwidth]{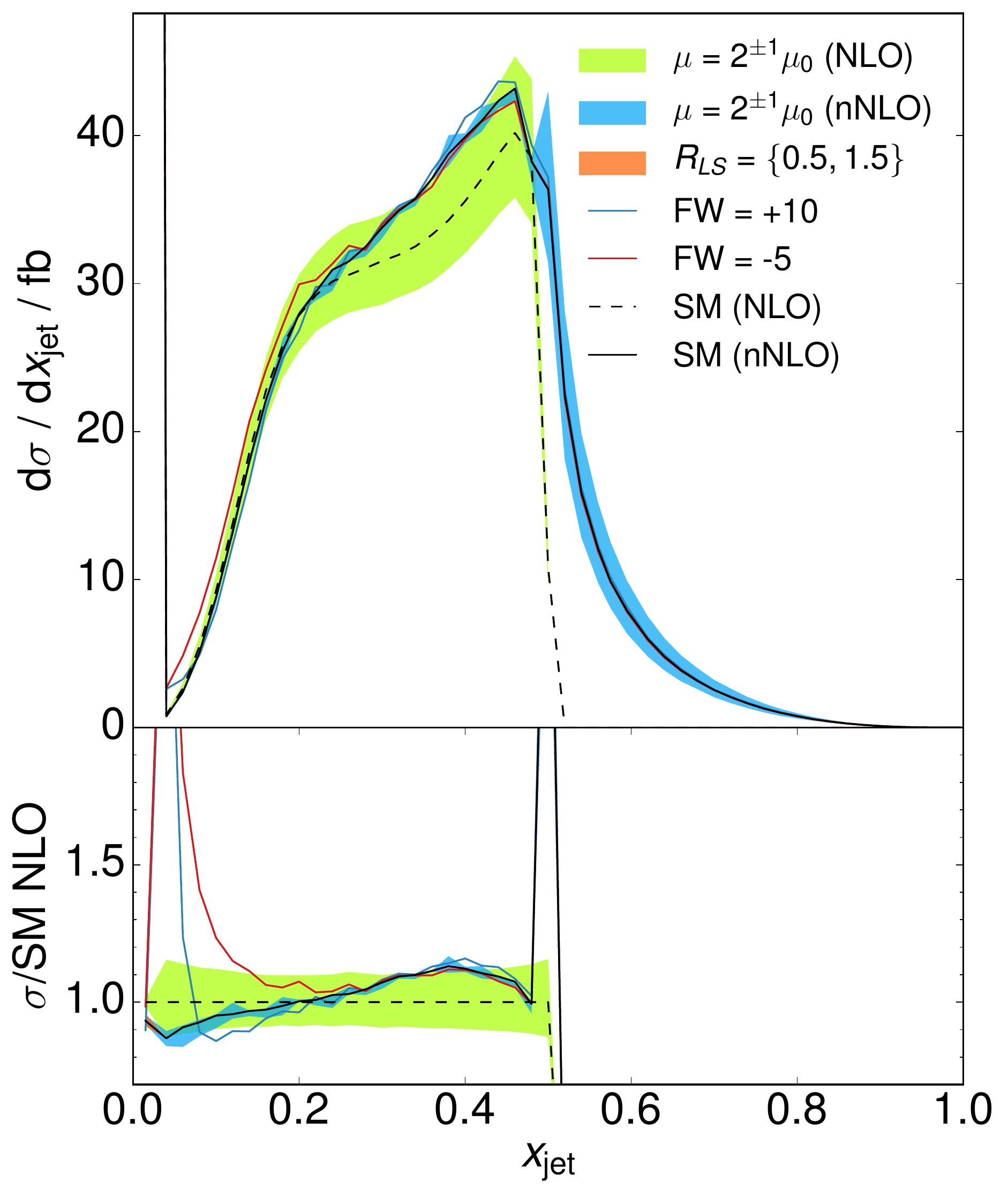}
    \includegraphics[width=0.45\textwidth]{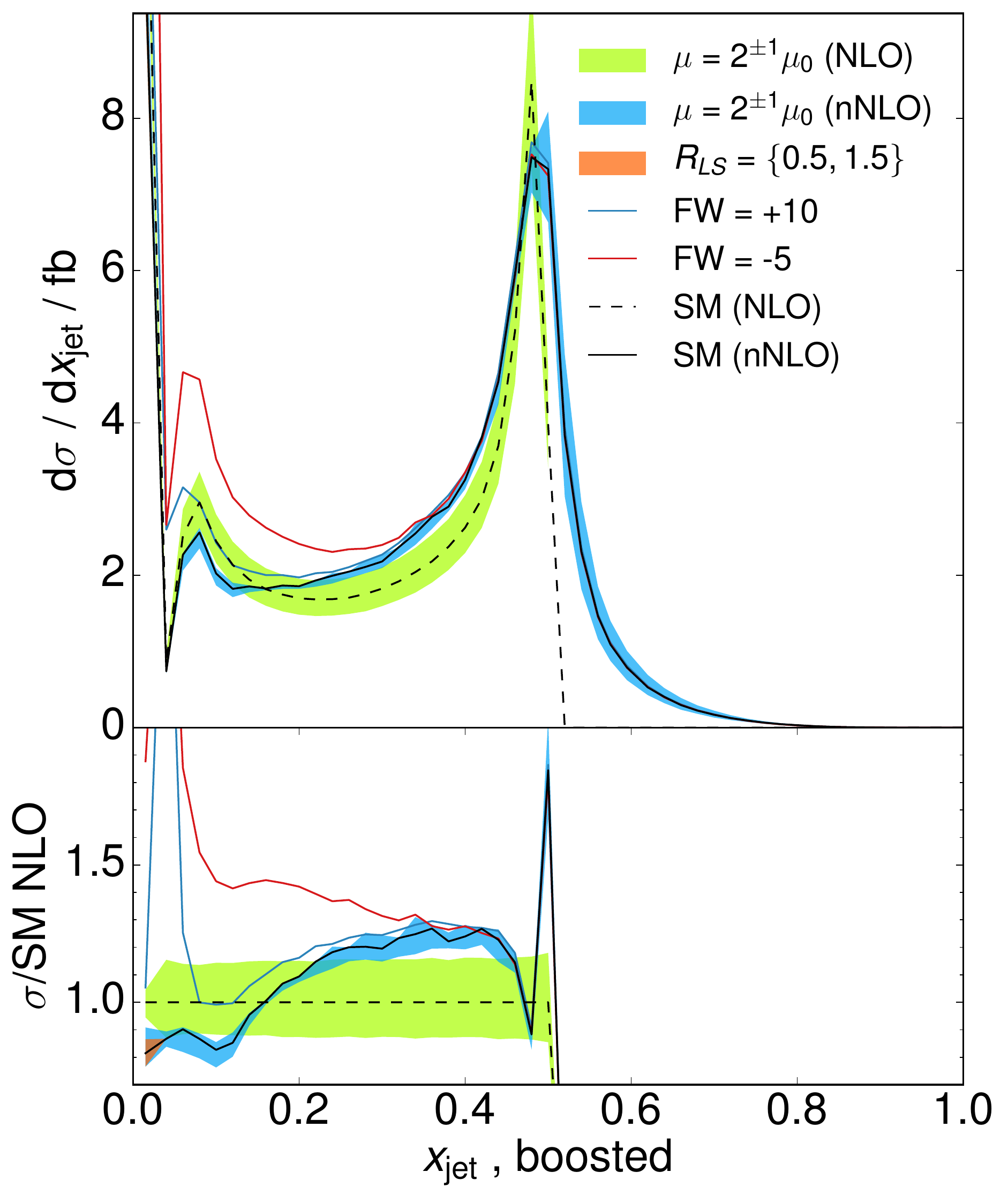}
  \end{center}
  \caption{
		$\xjet$ as defined in \cref{eq:xjetdef} for inclusive and boosted ($\ptz > \SI{200}{GeV})$ cuts.
		As visible especially in the boosted case, there are two relevant phase space regions
		around small $\xjet$ and for $\xjet \approx 0.5$.
		The latter is not sensitive to AC, such that the AC analysis should focus on the region below
		$\xjet = 0.2~\text{to}~0.3$.
  }
  \label{fig:xjet-nnlo-ac}
\end{figure}

In \cref{fig:xjet-nnlo-ac}, we show the $\xjet$ distribution at NLO and \nNLO QCD for the SM
as well as for two exemplary AC values at \nNLO QCD.
The ratio to the SM NLO prediction shows a non-flat effect of the \nNLO corrections,
enhancing large $\xjet$ values.
AC contribute at small $\xjet$ values.
Vetoing $\xjet > 0.2$ cuts away both the region not sensitive to AC as well as the region
with the largest higher-order corrections.

The veto only induces modest logarithms proportional to $\ln(\xjet) = \ln(0.2)$  
and offers an alternative to a fixed-$\pt$ veto without the need for resummation. 
\nNLO is a good testing ground for jet vetos, as all terms with potentially large logarithms
are included.
Jet veto studies at NLO are not sufficient as they miss the two-jet final state, which is
dominant in some phase space regions.

\begin{figure}[p]
  \begin{center}
    \includegraphics[width=0.45\textwidth]{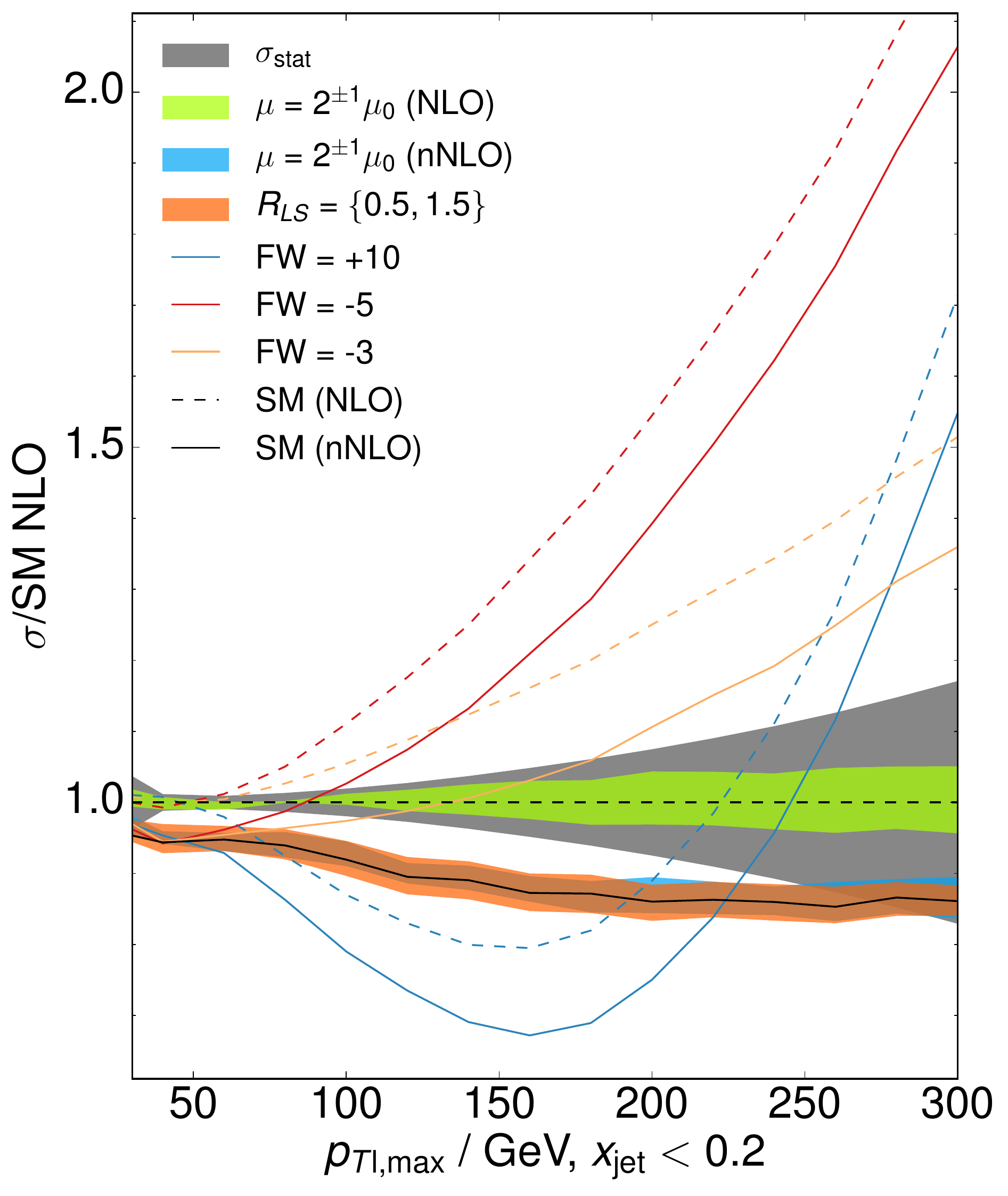}
    \includegraphics[width=0.45\textwidth]{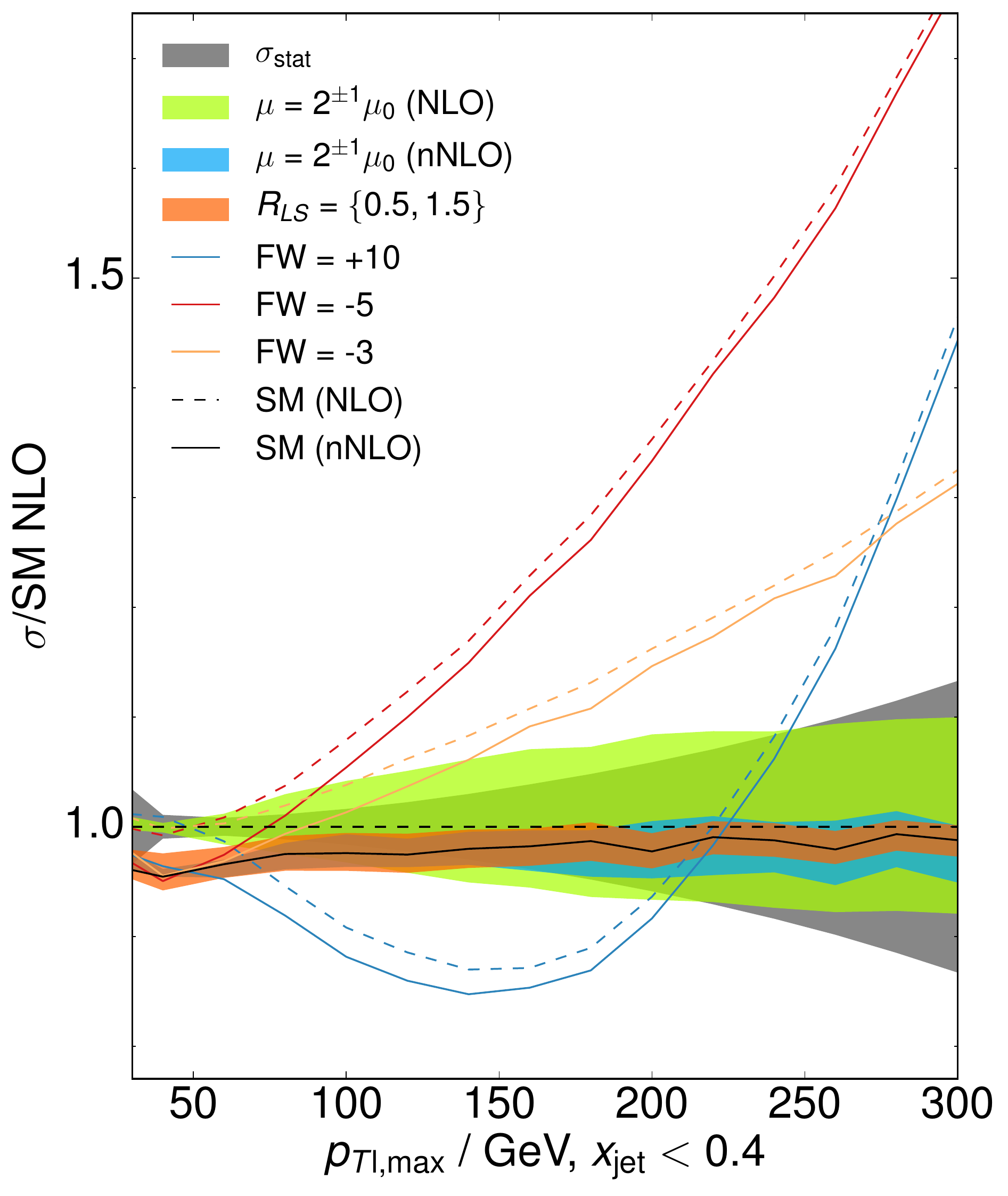}
  \end{center}
  \caption{
    The effect of anomalous couplings in combination with a jet veto is shown for the cuts
    $\xjet < 0.2$ and $\xjet < 0.4$. 
    Scale variation bands are given at NLO and \nNLO.
    An estimate of the statistical uncertainty is given as a grey band, for which
    an integrated luminosity of \SI{300}{fb^{-1}} is assumed as
    well as a factor $4$ for all lepton flavor combinations in the decays.
  }
  \label{fig:ptl-xjet-veto-ac}
\end{figure}

\Cref{fig:ptl-xjet-veto-ac} shows the $\pt$ of the hardest lepton when an additional cut
based on $\xjet$ is introduced.
The veto is designed to cut away jet-dominated events while allowing harder radiation than a
traditional fixed jet veto would, especially in the tails of distributions.
The veto reduces the \nNLO corrections. Instead of an increase of 25\%, we see a decrease of
the cross section by up to 10\%.
By cutting away the hard jet events, it also increases the sensitivity to AC.

In \cref{fig:ptl-xjet-veto-ac} scale variation bands are given at NLO and \nNLO.
Those bands are not a reliable uncertainty estimate, since with jet vetoes the scale
dependence is artificially reduced.
For an extended discussion and possible better estimates see e.g.
Ref.~\cite{Stewart:2011cf}.  
Besides the scale variation bands, we also show a band for the assumed statistical error.
This is a rough estimate based on \SI{300}{fb^{-1}} of data in \SI{20}{GeV} bins.
For extrapolation from one combination of lepton families in the decays of $W$ and $Z$ bosons to all 4
possibilities, we assume a factor $4$. In final states such as $\mu^+ \nu_\mu \mu^+ \mu^-$, there
are two different combinations possible for the reconstruction of the $Z$ boson. 
We neglect corrections due to imperfect reconstruction and to identical particle effects in that case.

\begin{figure}[p]
  \centering
  \includegraphics[width=.5\textwidth]{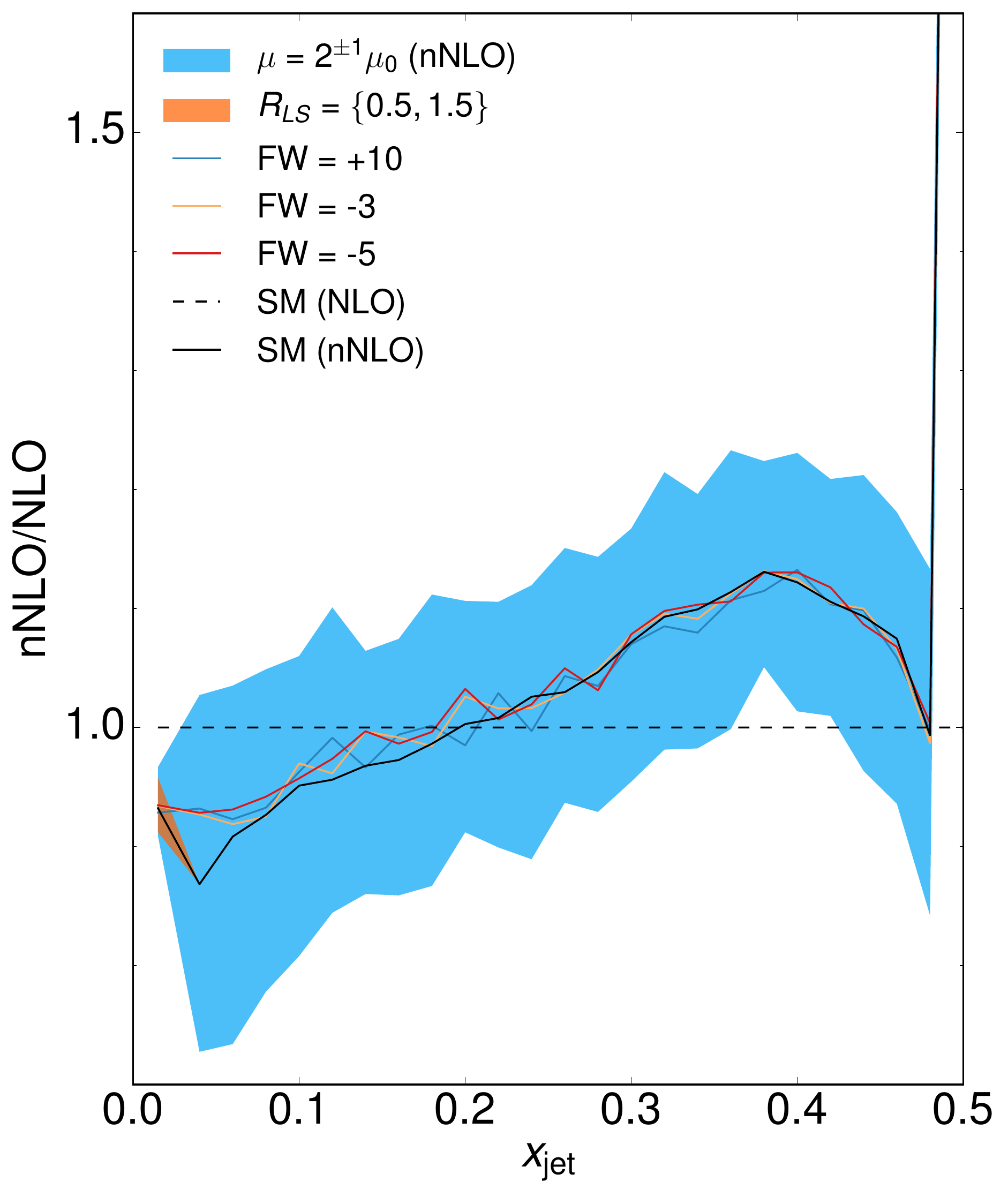}
  \caption{The K-factor for $\xjet$ of $\nNLO$ compared to NLO is shown for the SM and two
    values of the $\ow$ coupling. 
    The change of $\nNLO$ effects visible in \cref{fig:xjet-nnlo-ac} can be reduced to
    their dependence on $\xjet$.
    Different AC values show similar $\xjet$ K-factors and one could thus approximate
    $\nNLO$ effects by correcting with the shown K-factor.
    The fluctuations visible in this plot are due to Monte Carlo statistics.  }
  \label{fig:xj-nnlo-kfac}
\end{figure}

Considering $\nNLO$ corrections to different observables, we find that SM and AC receive
corrections with different shapes, such that a description of AC that is based on LO or NLO
and rescaled with SM K-factors is not sufficient.
In \cref{fig:xj-nnlo-kfac}, the \nNLO corrections to $\xjet$ are shown for the SM and
different AC values.
They are identical (within the simulation uncertainty), such that potentially a K-factor
binned in $\xjet$ could describe AC beyond NLO QCD.

\section{Conclusions}
\label{sec:conclusions}

We presented a calculation of WZ production at the LHC at $\nNLO$ QCD including Anomalous
Couplings (AC) using VBFNLO in combination with LoopSim.

To enhance the sensitivity to AC, we use a jet veto based on $\xjet$, as defined in \cref{eq:xjetdef}.
This reduces the large non-AC contribution where a high-$\pt$ jet recoils against a vector boson.
The $\xjet$ cut improves the sensitivity to anomalous couplings without introducing large
logarithms, as is expected for the traditional fixed-$\pt$ jet veto.
Furthermore, the $\xjet$ veto scales with the hardness of the event, such that we include
more phase space in the tails of the distribution.
Thereby the $\xjet$ cut preserves the region sensitive to AC also at high $\ptz$.

While currently most limits on AC are based on the high-$\pt$ tails of distributions, we
suggest to also study the interference region. Since high precision NNLO calculations for
vector boson pair production are now becoming available also for
distributions~\cite{Grazzini:2016ctr}, the theoretical uncertainties in the interference
regions will soon be small enough for a meaningful analysis. This double analysis has the
advantage of being sensitive to new physics from both strong coupling (large deviations in the high energy tail) as well as to intermediate coupling physics at lower energy scales, where also electroweak corrections are still expected to be modest.

The dominant kinematical effects of corrections are already present at NLO QCD. Thus, exploratory analyses for AC measurements can be performed with NLO programs. For full data analysis, however, the higher precision of NNLO calculations will ultimately be necessary.

\begin{acknowledgments}
\vspace{-1em}

FC has been partially supported by the Spanish Government and ERDF
funds from the European Commission (Grants No. FPA2014-53631-C2-1-P  
, FPA2014-57816-P, and SEV-2014-0398)
RR is supported by the Graduiertenkolleg ``GRK  1694:   Elementarteilchenphysik  bei
 höchster Energie und höchster Präzision.''
\end{acknowledgments}

%%%%%%%%%%%%%%%%%%%%%%%%%%%%%%%%%%%%%%%%%%%%%%%%%%%%%%%%%%%%%%
%\bibliographystyle{h-physrev} 
\bibliographystyle{JHEP} 
\bibliography{LSWZAC}

\end{document}